\documentclass[english]{article}
\usepackage[T1]{fontenc}
\usepackage[latin9]{inputenc}
\usepackage{esint}

\makeatletter

\usepackage{babel}
\makeatother

\begin{document}

\title{How often does theory match experiment?}

\author{\textbf{Anirban Banerji}\\
Bioinformatics Centre, University of Pune,\\
Pune-411007, India\\
anirbanab@gmail.com, anirban@bioinfo.ernet.in}

\maketitle
\begin{abstract}
In every sphere of science, theories make predictions and experiments
validate them. However, common experience suggests that theoretically
predicted exact magnitude for a parameter, constitute a small subset
of all the experimentally obtained magnitudes for that particular
parameter. Typically, irrespective of the branch of science and the
particular problem under consideration, the set of obtained experimental
results form an interval $[x_{min},x_{max}]$, within which the theoretically
predicted magnitude, say $x$, occurs with time, apparently randomly.
We attempt here to find the characteristics of the statistical distribution
of events of experimental observation of the occurrence of theoretically
predicted $x$; in other words, characterization of the time interval
when theoretical predictions match the experimental readings, exactly.\\

\end{abstract}
Recording the readings from experimental apparatus is ubiquitous in
every branch of science. While the theoretical studies help us to
predict the expected magnitude that any parameter should own, experiments
verify whether the parameters actually assume the predicted magnitudes;
and if they don't, by what margin do they differ from the theoretical
idealizations. In some of these cases experiments are one-off in nature;
whereas in many cases, we gather results from the apparatus in a sequential
manner before studying the extent by which the mean and variance of
the experimentally obtained data differs from the theoretical predictions
(if they do, at all). In case of the later, we collect the results
as a sequence of homogeneous events occurring one after another, which
implies that the occurrence of results constitute a flow of events.
There might or might not be a fixed time interval between the occurrences
of these results. In the present work we attempt to understand the
nature of the process of observing the occurrences of experimental
results, for the general case where they occur without any fixed time
interval between them. Since we know it from our experience that the
precise magnitude of theoretically predicted result (from a pool of
results that differ from the exact expectation by arbitrarily small
margin) shows up in the apparatus only now and then and not always;
attempts are made to characterize the time intervals between occurrences
of results, when theoretical idealizations match absolutely with the
experimental realities (referred to as 'events' from here onwards).\\
\\
It is not that attempts have not been made to tackle this problem.
On the contrary, previous attempts to characterize this problem were
many (and they form a spectrum of standpoints); but most of them were
qualitative in nature and were only tangentially touching upon the
mathematical description of the situation. For example, {[}1] talks
about certain prevalent patterns in experimental observation in particular
cases related to states in Alzheimer\textquoteright{}s disease whereas
{[}2], {[}3] and {[}4] had attempted to understand the general philosophical
nature of the problem from various perspectives. Although {[}5 ] and
{[}6] were objective and quantitative in their basic premise of description,
the precise question that we are raising in this work was not addressed.
Similarly, although the attempts of {[}7], {[}8] and {[}9] had touched
upon the extent to which theoretical predictions matched experimental
findings in various experimental cases, the statistical characterization
of precise time-intervals between two events where theoretically predicted
magnitude for a parameter matches exactly with experimentally obtained
values, remained unanswered. \\
\\
\\
Experience suggests that results are generally produced one at a time,
for any arbitrarily small time interval, and not in a group of two,
three per time interval; for most of the experiments across the spectrum
of scientific streams. This implies that the probability of two or
more events occurring in an elementary time interval $\triangle t$
is negligibly small compared to the probability of single events occurring
at different time intervals(arbitrarily chosen); and hence we conclude
that the flow of events (when theoretically predicted result matches
absolutely with experimentally obtained results) is ordinary{[}10].
We observe further that the probability characteristic describing
the nature of occurrences of these results do not depend upon the
choice of some particular reference frame; i.e., the flow of events
is stationary{[}11]. Experience teaches us further that the number
of events occurring on any particular time interval, say $t_{1}$,
does not, in general, depend upon number of events occurring on any
other non-overlapping interval. These characteristics imply that the
flow of events (occurrences of results in a sequential manner) can
be represented as an stationary Poisson flow{[}12]. Alongside all
of these, a closer look of the problem reveals that in most of the
cases, the time intervals $T_{1},T_{2},\ldots$ between successive
events are independent. We assume, with no loss of generality, that
intervals between $T_{1},T_{2},\ldots$ are similarly distributed
random variables. Aforementioned observations and the assumptions
tend to suggest that flow of occurrences of results when theoretical
prediction matches experimentally obtained results absolutely, can
be categorized as a 'recurrent flow' with limited after-effects {[}13]
(in other words, a 'Palm flow').{[}14]\\
\\
\\
We now attempt to describe an extremely familiar situation when an
experimentalist is taking the readings to verify his/her prediction
(from theoretical studies) about what to expect in terms of exact
magnitude of some parameter. In many of the situations like this,
the desired events of occurrence of predicted magnitude occurs, but
only as an element of a set of produced results, where most of the
elements of the set differ from the expected magnitude by a little.
Any experimentalist, from any branch of science, is aware of such
a situation; but knowledge of any pattern in the occurrences of aforementioned
desired results, eludes the students of science.\\
\\
To describe the situation, we start by assuming that the occurrences
of the theoretically predicted result constitute a 'recurrent flow'
on the time axis and the intervals between any two occurrences of
them, follow a distribution with density $f(t)$. The assumption that
the pattern of output generation can be described by $f(t)$ is not
unrealistic; because, results occur as an output of a definite, deterministic
process (carefully planned experiment, in any branch of science).\\
\\
We denote the exact moments of occurrence of the events of absolute
match between theoretically predicted results and experimentally found
ones, by $\tau$. We seek to find the distribution density of $f^{\tau}\left(\tau\right)$
of the interval $T^{\tau}$, where the points in the time axis $\tau$
occur. This implies, calculation of probability where $f^{\tau}\left(\tau\right)dt$
equals that of $\tau$'s occurrence between length of time interval
$\left(t,t+\triangle t\right)$. Assuming the presence of very large
number of intervals (N) between the events constituting the entire
temporal extent of the experiment $\Gamma$; we find that the average
number of intervals with length in the range $\left(t,t+\triangle t\right)$
is $Nf(t)dt$; whereas the average total length of all such intervals
equals $tNf(t)dt$. The average total length of all the N intervals
on $\Gamma$ can be represented as $Nx_{t}$, where $x_{t}$ denotes
the expectation $E\left[T\right]=\intop_{0}^{\infty}tf(t)dt$. Hence
:\\
\begin{equation}
f^{\tau}\left(\tau\right)dt\approx\frac{tNf(t)dt}{Nx_{t}}=\frac{tf(t)}{x_{t}}dt\end{equation}
\\
The approximation in eq$^{n}$1 becomes more exact when longer interval
of time $\Gamma$ is considered (larger N). Distribution of the random
variable $T^{\tau}$ can then be found by evaluating the limit,\\
\begin{equation}
f^{\tau}\left(\tau\right)=\frac{t}{x_{t}}f\left(t\right)\left(t>0\right).\end{equation}
\\
\begin{equation}
E\left[T^{\tau}\right]=\frac{1}{x_{t}}\int_{0}^{\infty}t^{2}f(t)dt=\frac{1}{x_{t}}\mu_{2}(t)=\frac{1}{x_{t}}\left(\sigma_{t}^{2}+x_{t}^{2}\right)\end{equation}
\\
\begin{equation}
\sigma^{2}\left[T^{\tau}\right]=\mu_{2}\left[T^{\tau}\right]-\left(E\left[T^{\tau}\right]\right)^{2}=\frac{1}{x_{t}}\int_{0}^{\infty}t^{3}f(t)dt-\left(E\left[T^{\tau}\right]\right)^{2}\end{equation}
\\
However, from an experimentalist's point of view, it would be more
useful to know the characteristics of the case where any time interval
$T^{\tau}$ is divided in two intervals $I_{1}$ and $I_{2}$, by
the occurrence of the events when theoretically predicted magnitude
for some parameter matches with the experimental findings, at a random
instance $\tau$. Here $I_{1}$ is defined by nearest previous event
to $\tau$ and $I_{2}$ is defined by the occurrence of $\tau$ to
the nearest successive event. Characterization of such a case will
be of immense practical help to the experimentalists.\\
\\
We approach the situation by assuming $T^{\tau}=\theta$. By introducing
a density $f_{I_{1}}\left(t|\theta\right)$ that describes the conditional
probability of the interval $I_{1}$ in the presence of $\theta$.
We observe that occurrence of the event of exact match between from
theoretical prediction and experimentally obtained result, is random
in time; and hence we can consider its having a uniform distribution
in the interval $\theta$, given by :\\
\begin{equation}
[f_{I_{1}}\left(t|\theta\right)=\frac{1}{\theta}],\forall0<t<\theta\end{equation}
\\
However, to find the marginal distribution $f_{I_{1}}\left(t\right)$,
we average the density(eq$^{n}$-5) considering, the weight $f^{\tau}\left(\tau\right)$.
Applying eq$^{n}$-2, we obtain :\\
\\
$f^{\tau}\left(\theta\right)=\frac{\theta}{x_{t}}f\left(\theta\right)$
and $f_{I_{1}}\left(t\right)=\int_{0}^{\infty}f_{I_{1}}\left(t|\theta\right)f^{\tau}\left(\theta\right)d\theta$\\
\\
But, since $f_{I_{1}}\left(t|\theta\right)$ is nonzero only for $\theta>t$,
we can write \\
\\
\begin{equation}
f_{I_{1}}\left(t\right)=\int_{t}^{\infty}\frac{\theta}{\theta x_{t}}f\left(\theta\right)d\theta=\frac{1}{x_{t}}\int_{t}^{\infty}f(t)dt=\frac{1}{x_{t}}\left[1-\Phi\left(t\right)\right]\end{equation}
\\
where $\Phi\left(t\right)$ is the distribution function of the interval
t between the events in the 'recurrent flow'. \\
\\
It is evident that $I_{2}$ $\left(I_{2}=T^{\tau}-I_{1}\right)$,
will have the identical distribution :\\
 \begin{equation}
f_{I_{2}}\left(t\right)=\frac{1}{x_{t}}\left[1-\Phi\left(t\right)\right]\end{equation}
\\
\textbf{\underbar{Conclusion :}}\\
With the help of this simple model a construct is proposed that could
characterize the mathematical nature of distribution of instances
when theoretical prediction about the magnitude of any parameter matches
with the experimental results. This distribution is found to resemble
the characteristics of a 'recurrent flow' with limited after-effects.
Since the number of assumptions involved in constructing this model
are kept at a minimal and the possible domain of applicability of
the aforementioned finding encompasses the entire gamut of scientific
paradigms (wherever the magnitude of any theoretically predicted parameter
is compared with mean and variance of the experimentally obtained
results), the model will hopefully serve students of science, across
the barriers of scientific streams.\\
\\
\textbf{\underbar{Acknowledgment :}} This work was supported by COE-DBT(Department
of Biotechnology, Govt. of India) scheme.\\
The author would like to thank the Director of Bioinformatics Centre,
University of Pune; Dr. Urmila Kulkarni-Kale and Professor Indira
Ghosh(his PhD supervisor), for providing him with the opportunity
to perform this work, although it has got nothing to do with his PhD.
research.\\
\\
\textbf{\underbar{References :}}\\
{[}1] Behl C., Oxidative stress in Alzheimer\textquoteright{}s Disease:
Implications for Prevention and Therapy, {}``Alzheimer\textquoteright{}s
Disease : Cellular and Molecular Aspects of Amyloid $\beta$ {}``,
Harris JR, Fahrenholz F (Eds.); Springer, USA; 2005, pp 65-78.\\
{[}2] Wuketits F.M. , Functional Realism, {}``Functional Models of
Cognition: Self-organizing Dynamics and Semantic Structures in Cognitive
Systems'', Arturo Carsetti (Eds), Springer, 1999, pp 27-38.\\
{[}3] McGrath A.E., {}``Science and Religion: An Introduction'',
Blackwell Pub, 1999; pp-78.\\
{[}4] Adams G. and Stocks E.L., A Cultural Analysis of the Experiment
and an Experimental Analysis of Culture, Social and Personality Psychology
Compass, 2 (5), 1895-1912.\\
{[}5] Vasileva L.A. and Gortsev A.M., Estimation of Parameters of
a Double-Stochastic Flow of Events under Conditions of Its Incomplete
Observability, Automation and Remote Control, 63(3), 2002, 511\textendash{}515.\\
{[}6] Arita M., Introduction to the ARM Database: Database on Chemical
Transformations in Metabolism for Tracing Pathways, {}``Metabolomics
: The Frontier of Systems Biology'', Tomita M. and Nishioka T., (Eds),
pp 193-210.\\
{[}7] Downs A.J., Himmel H.J and Manceron L., Low valent and would-be
multiply bonded derivatives of the Group 13 metals Al, Ga and In revealed
through matrix isolation, Polyhedron, 21(5-6), 2002, 473-488.\\
{[}8] Judd B.R, Complex atomic spectra, 1985, Rep. Prog. Phys., 48,
907-954.\\
{[}9] Ramirez-Duverger A.S., and Llmas R.G., Light scattering from
a multimode waveguide of planar metallic walls, Optics Communications,
227(4-6), 2003, 227-235.\\
{[}10] Kovalenko I.N., Kuznetsov N.Y. and Shurenkov V.M.; Models of
Random Processes: A Handbook for Mathematicians and Engineers; 1996;
Boca Raton: CRC Press, pp.155\\
{[}11] Ibid. pp-162.\\
{[}12] Ibid. pp-77.\\
{[}13] Kalashnikov, V., Mathematical Methods in Queuing Theory, Springer,
1994,pp-10.\\
{[}14] Kovalenko I.N., Kuznetsov N.Yu. and Shurenkov V.M.; Models
of Random Processes: A Handbook for Mathematicians and Engineers;
1996; Boca Raton: CRC Press, pp.161.
\end{document}